\documentclass[
twocolumn,
superscriptaddress,
showpacs,
amsmath,amssymb,
aps,
prl,
]{revtex4-1}

\pdfoutput=1

\usepackage{graphicx}
\usepackage{dcolumn}
\usepackage{bm}

\begin{document}

\title{ Study of the $5p_{3/2} \rightarrow 6p_{3/2} $ electric dipole forbidden transition in atomic rubidium using optical-optical double resonance spectroscopy.}


\author{F. Ponciano-Ojeda}
\author{S. Hern\'andez-G\'omez}
\author{O. L\'opez-Hern\'andez}
\author{C. Mojica-Casique}
\author{R. Col\'\i n-Rodr\'\i guez}
\author{F. Ram\'irez-Mart\'inez}\email[]{ferama@nucleares.unam.mx}
\author{J. Flores-Mijangos}
\affiliation{Instituto de Ciencias Nucleares, UNAM. Circuito Exterior, Ciudad Universitaria, 04510 M\'exico. D.F., M\'exico.}

\author{D. Sahag\'un}
\author{R. J\'auregui}
\affiliation{Instituto de F\'\i sica, UNAM, Ciudad Universitaria, 04510 M\'exico D.F., M\'exico.}

\author{J. Jim\'enez-Mier}\email[]{jimenez@nucleares.unam.mx}
\affiliation{Instituto de Ciencias Nucleares, UNAM. Circuito Exterior, Ciudad Universitaria, 04510 M\'exico. D.F., M\'exico.}

\date{\today}

\begin{abstract}
Direct evidence of excitation of the $5p_{3/2} \rightarrow 6p_{3/2} $ electric dipole forbidden transition in atomic rubidium is presented.
The experiments were performed in a room temperature rubidium cell with continuous wave extended cavity diode lasers.
Optical-optical double resonance spectroscopy with counterpropagating beams allows the detection of the non-dipole transition free of Doppler broadening.
The $5p_{3/2} $ state is prepared by excitation with a laser locked to the maximum $F $ cyclic transition of the D2 line, and the forbidden transition is produced by excitation with a $911 $ nm laser.
Production of the forbidden transition is monitored by detection of the $420 $ nm fluorescence that results from decay of the $6p_{3/2} $ state.
Spectra with three narrow lines ($\approx 13 $ MHz FWHM) with the characteristic $F-1 $, $F $ and $F+1 $  splitting of the $6p_{3/2} $ hyperfine structure in both rubidium isotopes were obtained.  
The results are in very good agreement with a direct calculation that takes into account the $5s \rightarrow 5p_{3/2} $ preparation dynamics, the $5p_{3/2} \rightarrow 6p_{3/2} $ non-dipole excitation geometry and the $6p_{3/2} \rightarrow 5s_{1/2} $ decay. 
The comparison also shows that the electric dipole forbidden transition is a very sensitive probe of the preparation dynamics. 
\end{abstract}

\pacs{32.70.Cs,32.70.Fw}
\maketitle


While the electric dipole approximation is a cornerstone in the study of the interaction between optical radiation fields and atoms, transitions induced by optical fields beyond this approximation have also become important tools in basic and applied studies of atoms.
These so called ``forbidden transitions'' have been traditionally used in astrophysical and plasma studies \cite{Biemont1996}. 
They now play a fundamental role in metrology \cite{Rooij2011} and have also been used in experiments testing parity non-conserving interactions in atoms \cite{Bouchiat1997}.    

In early studies of forbidden transitions, \citet{Sayer1971}  determined transition probabilities of electric quadrupole (E2) transitions using a tungsten lamp. The first direct observation of electric quadrupole effects in multiphoton ionization dates back to the work of \citet{Lambropoulos1975}. Electric-dipole-forbidden transitions were exploited in three-wave-mixing experiments for optical sum and difference frequency generation in~\citep{Flusberg1977a,*Flusberg1977b}. 

The use of intense continuous-wave or pulsed laser sources has facilitated the observation of weak absorption lines. For instance, \citet{Tojo2005} reported a determination of the oscillator strength of a E2 transition with a temperature-controlled cell and an extended cavity diode laser. Also, the study of strongly forbidden $J=0 \to J=0$ transitions via single-photon excitation is presented in \citep{Taichenachev2006,*Barber2006}. Excitation of forbidden transitions involving states with nonzero angular momentum in alkali atoms have also been studied over the last few years~\cite{Mironova2005a,*Mironova2005b,Bayram2000,Bhattacharya2003,Pires2009,Tong2009}. The coherent mixing of waves is theoretically studied in \citep{Mironova2005a,*Mironova2005b} for $n_1\; ^2P- n_2\; ^2P$ transitions. The excitation of the $5p \rightarrow 8p $ forbidden transition in thermal rubidium atoms was produced with a pulsed laser in \cite{Bayram2000} and using cold atoms in \cite{Pires2009}.
The experiment with cold atoms \cite{Pires2009} allowed resolution of the atomic hyperfine structure and conclusively determined that there was no magnetic dipole contribution to this transition.
Other experiments with dipole forbidden transitions and cold alkali atoms include the measurement of the $3p \rightarrow 4p $ transition in sodium \cite{Bhattacharya2003} and also the $5s \rightarrow nd $ transitions in rubidium \cite{Tong2009}. Recently, experiments performed in atomic vapor nano-cells with a half-wavelength thickness and an applied magnetic field demonstrated a strong enhancement of the probabilities of forbidden transitions \cite{Hakhumyan2012,*Sargsyan2014}.


In this article we present experimental results for the excitation of the $5p_{3/2} \rightarrow 6p_{3/2} $ electric dipole forbidden transition in atomic rubidium.
The experiment was performed with thermal atoms and continuous wave extended cavity diode lasers (ECDL), and we were able to resolve the hyperfine structure of the $6p_{3/2} $ state.
As far as we know this is the first time that a non-dipole experiment with such a high resolution has been performed with room temperature rubidium atoms.
Our results clearly indicate that similar experiments can be performed with the other alkali atoms.
This is in agreement with the observation in \cite{Bhattacharya2003} that moderate cw laser powers could be used to excite the $np \rightarrow (n+1)p $ forbidden transition in any of the alkalis. 
The data are compared with the results of a calculation that considers three independent steps, namely, state preparation, non-dipole excitation and decay of the $6p_{3/2} $ levels, where the first and third steps are dipole transitions whereas the non-dipole excitation is an electric quadrupole transition. 

\begin{figure}
   \centering
       \includegraphics[width=0.9\linewidth]{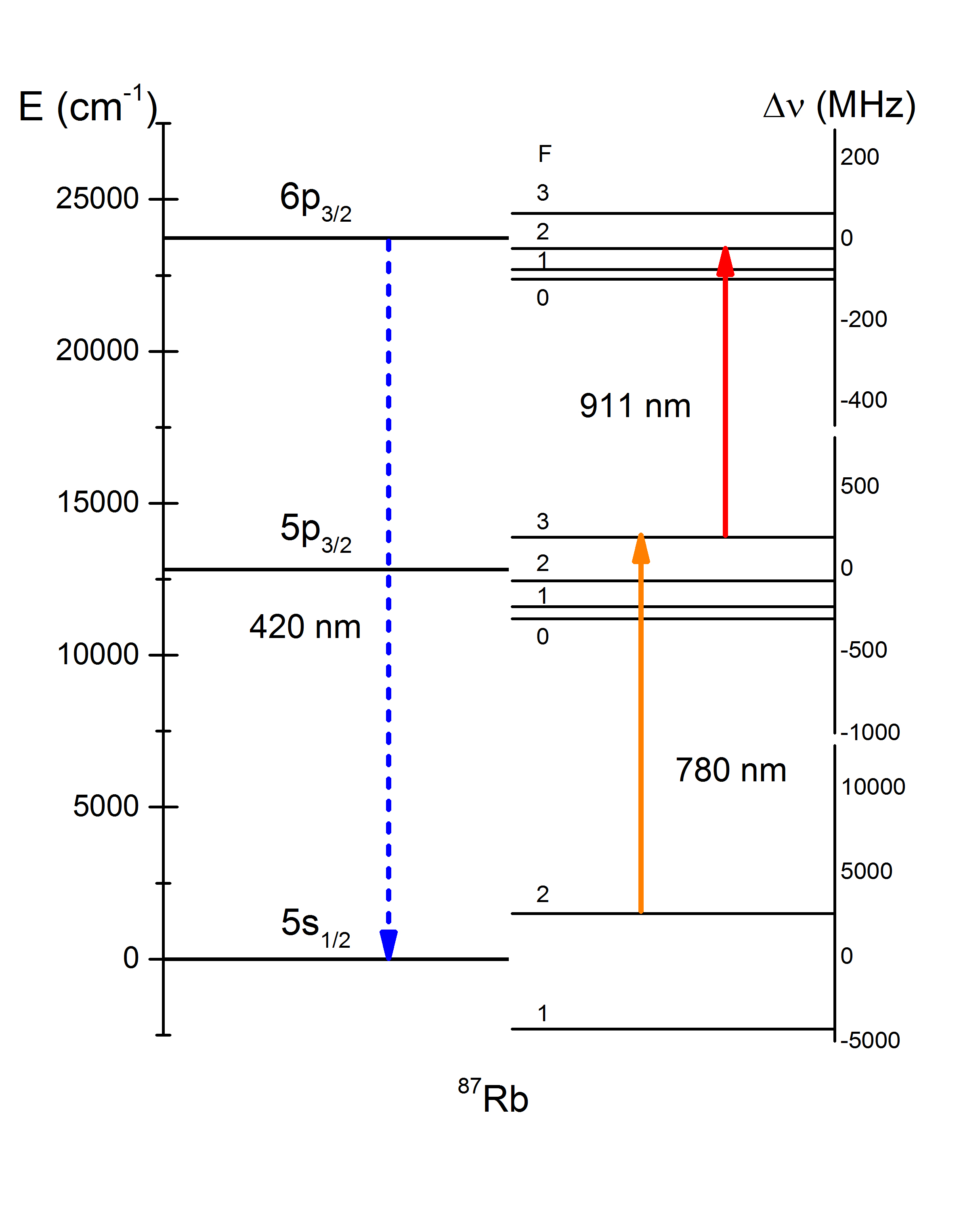}
       \caption{\label{fig:Rbenlvl} Energy levels of $^{87}$Rb. The left panel includes the fine structure. The hyperfine structure is shown to the right. Note that the frequency scale changes for the hyperfine structure of each state.}
\end{figure}

Figure \ref{fig:Rbenlvl} shows an energy level diagram, where the total angular momentum quantum numbers and the hyperfine splittings correspond to $^{87} $Rb. 
A similar figure, with different values of $F $ and hyperfine splittings, is obtained for $^{85}$Rb.
In the experiment a  laser in resonance with the $5s \rightarrow 5p_{3/2} $ transition at $780 $ nm (D2 line) is used to prepare atoms in the $5p_{3/2} $ state.
A second laser beam at $911 $ nm is used to produce the $5p_{3/2} \rightarrow 6p_{3/2} $ electric dipole forbidden transition. 
We detect this excitation channel because atoms in the $6p_{3/2} $ state have a significant probability of decaying directly into the $5s $ ground state by emission of a $420 $ nm photon.

\begin{figure}
   \centering
       \includegraphics[width=\linewidth]{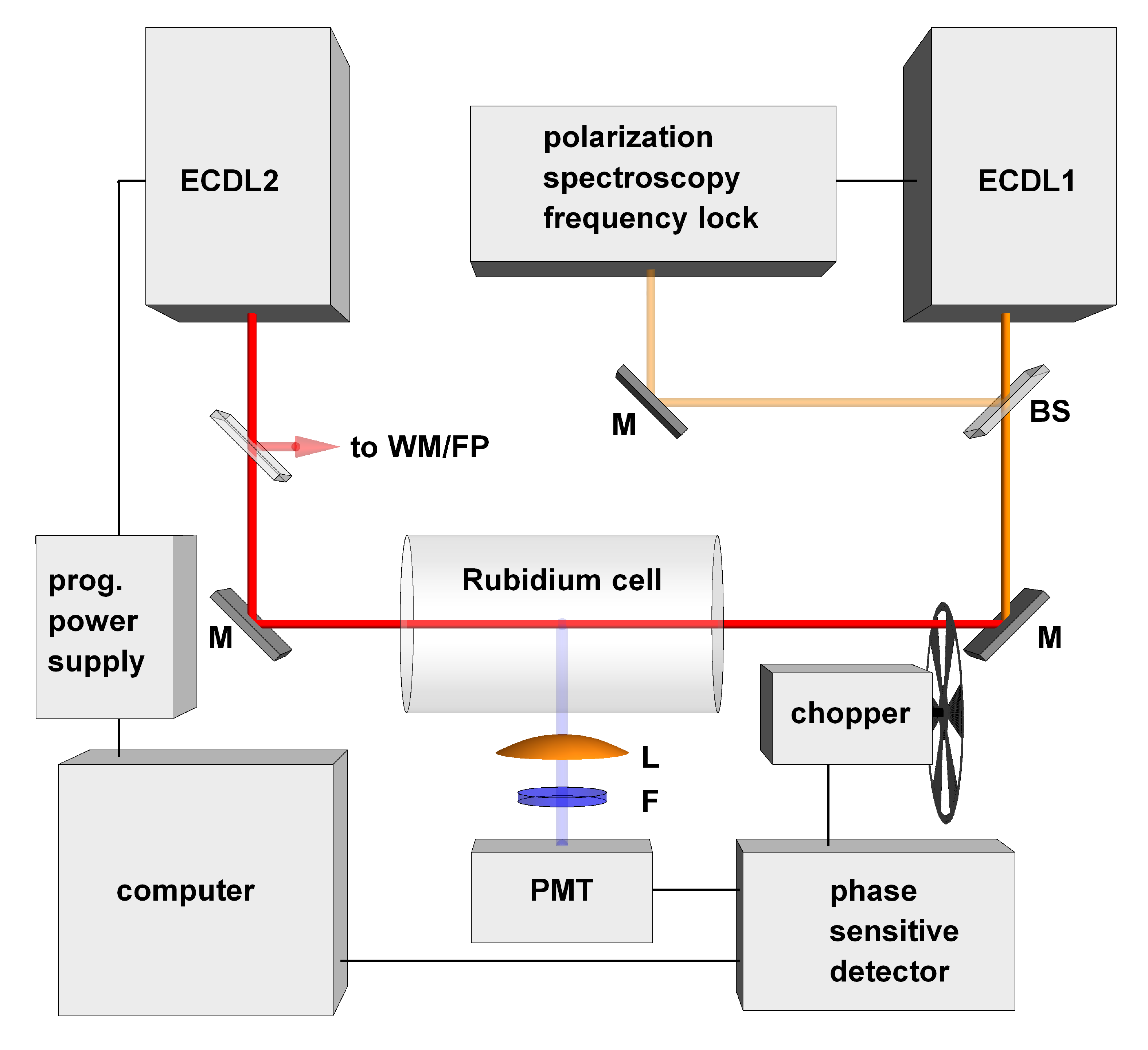}
       \caption{\label{fig:setup} Experimental setup for polarized
       velocity selective spectroscopy. ECDL: extended cavity diode laser; PMT: photomultiplier tube; M: mirror, BS: beamsplitter, L: lens system; F: $420 $ nm interference filter. A small part of the $911 $ nm beam is sent to a wavemeter (WM) and a $1.5 $ GHz Fabry-Perot interferometer (FP).}
\end{figure}

The experimental setup is shown in Fig. \ref{fig:setup}.
Home made extended cavity diode lasers (ECDL) provide the $780 $ and $911 $ nm photon beams.
Both lasers were built after the design of refs. \cite{Arnold1998,Hawthorn2001}, adapted to the emission wavelengths.
ECDL1 operates at the frequency of the D2 transition in atomic rubidium ($780 $ nm).
It has an emission bandwidth of less than $ 6 $ MHz.
Its frequency can be locked to the Doppler-free cyclic transition of either of the rubidium isotopes \cite{Pearman2002,Harris2006}.
ECDL2 uses a laser diode with a nominal emission wavelength of $915 $ nm \cite{Axcel}.
In the extended cavity configuration its emission can be tuned to $911 $ nm.
It operates in a single mode and can be tuned across mode-hop free regions of $\approx 3 $ GHz.
Under normal operation conditions one obtains up to $100 $ mW of single-mode laser power. 
Both beams are linearly polarized with parallel electric field vectors, and counterpropagate along a rubidium cell at room temperature.
The production of the electric dipole forbidden transition is monitored by detecting the $420 $ nm fluorescence that results from the direct decay into the $5s $ ground state.
These blue photons are collected by a lens system that focuses them into the cathode of a photomultiplier tube (PMT).
A bandpass filter centered at $420 $ nm is placed in front of the PMT window.
In the experimental setup the detection direction is perpendicular to the linear polarization direction of both laser beams, and also perpendicular to their propagation direction.
A chopper is used to modulate the incidence of the preparation beam with a frequency of $800 $ Hz.
The amplified PMT current signal and the chopper reference frequency signal are sent to a phase sensitive detector whose voltage output is read in a computer interface.
This interface also controls a programable power supply that is used to scan the frequency of ECDL2.
An electric dipole forbidden spectrum is the in-phase PMT signal as a function of the voltage applied to the frequency scan of ECDL2.
A wavemeter with a $50 $ pm resolution is used for the initial tuning of ECDL2.
A confocal Fabry-Perot interferometer ($1.5 $ GHz FSR) is used to monitor the single-mode operation of ECDL2, and it also provides a coarse frequency scale.

Both laser beams were collimated and produced elliptical cylinders along the $7.5 $ cm long rubidium cell.
The ECDL1 beam profile was a $4.9 $ mm $\times 2.4 $ mm ellipse and that of ECDL2 was a  $4.5 $ mm $\times 2.3 $ mm ellipse, respectively, with almost complete overlap inside the rubidium absorption cell.
For the $911 $ nm beam we used the maximum available power of $100 $ mW, which results in an average intensity of $12.3 $ kWm$^{-2}$.  
The fluorescence lines can be broadened by the power of the $780 $ nm preparation beam.
We therefore decided to use $100 \ \mu$W of power.
This puts its average intensity at $10.7 $ Wm$^{-2}$, below the $16.46 $ Wm$^{-2}$ saturation intensity for the D2 transition \cite{Metcalf1999}.


In either rubidium isotope, the $5s_{1/2} $ hyperfine splitting is larger than the D2 Doppler width at room temperature. 
Therefore, the frequency of the preparation photons at $780 $ nm select the initial hyperfine state of the three-step excitation sequence.
We used polarization spectroscopy \cite{Pearman2002,Harris2006} to lock the frequency of the preparation beam to the Doppler free $F \rightarrow F+1 $ cyclic transition ($F = 2$ in $^{87} $Rb or $F=3 $ in $^{85} $Rb).
The $911 $ nm laser is used to excite the $5p_{3/2} \rightarrow 6p_{3/2} $ electric dipole forbidden transition.
By sending it in a counterpropagating configuration one can perform a Doppler free excitation into the hyperfine states of the $6p_{3/2} $ manifold  \cite{Kaminsky1976}.
For zero velocity atoms the excitation sequence is $F_1=F \rightarrow F_2=F+1 \rightarrow F_3 $.
Direct use of the electric quadrupole selection rules ($\Delta F = 0$, $ \pm 1$, $ \pm 2$) results in $  F_3 = 1$, $2 $ and $3 $ for $^{87}Rb $ and  $F_3 = 2$, $3 $ and $4$ for $^{85} $Rb. 
For each isotope one therefore expects a triplet with the frequency splitting of the well known hyperfine structure of the $6p_{3/2} $ state \cite{Arimondo1977}.
These splittings were used for the frequency calibration of the dipole-forbidden spectra.

The relative intensities of the emission that follows the electric dipole forbidden excitation are also calculated.
Assuming three sequential steps (preparation, electric quadrupole excitation, and decay), the probability to observe a $420 $ nm photon resulting from decay of the $\left|6p_{3/2} F_3 \right\rangle $ hyperfine states is given by:
\begin{widetext}
\begin{equation}
P(F_3)  = \sum_{M_2,M_3,F_1^\prime,M_1^\prime} \sigma(F_2,M_2) \left|\left\langle 6p_{3/2}  F_3 M_3 \left| T \right| 5p_{3/2} F_2 M_2 \right\rangle \right|^2    \left|\left\langle 5s_{1/2}  F_1^\prime M_1^\prime \left| D \right| 6p_{3/2} F_3 M_3 \right\rangle\right|^2
\label{eq:signal}
\end{equation}
\end{widetext}
\noindent Here $\sigma(F_2,M_2) $ is the population of the $\left|5p_{3/2} F_2 M_2 \right\rangle $ prepared by the $780 $ nm laser, $T $ is the non-dipole transition operator, and $D $ is the $6p_{3/2} \rightarrow 5s_{1/2} $ electric dipole decay operator. The sum is performed over all projections of total angular momenta of the initial  ($M_2 $) and final ($M_3 $) states of the non-dipole transition, and also over the angular momenta of the final $5s_{1/2} $ hyperfine states $(F_1^\prime,M_1^\prime)$.
The value of the total angular momentum of the intermediate state $F_2 $ corresponds to the $F \rightarrow F+1 $ cyclic transition of the D2 preparation step.  
This expression assumes that the forbidden excitation is very weak compared to the preparation step.
Therefore, both electric quadrupole excitation and its subsequent electric dipole decay do not modify the populations of the $5p_{3/2} $ state.
However, optical pumping in the preparation step plays a very important role in establishing the populations $\sigma(F_2,M_2) $ \cite{Harris2006}.
These populations were calculated for linearly polarized light using the rate equation approximation and taking into account the transit time of the atoms across the preparation beam \cite{Harris2006}. 
In a study of the $5p \rightarrow 8p $ dipole forbidden transition in cold rubidium atoms \cite{Pires2009}, no significant contribution of magnetic dipole transitions was found, and there is no reason why it should appear in our experiment.
Therefore, in the present calculation we only used an electric quadrupole transition operator for $T $.
In our experimental geometry, with the z-axis along the linear polarization of both preparation and excitation lasers, and taking the propagation direction along the x-axis, the non-dipole transition element is $T \propto xz $.
Finally, the $6p \rightarrow 5p $ decay is observed along the y-axis, with no polarization selection. 
Therefore, for decay we took incoherent sums of the $D = x $ and $D = z $ electric dipole operators.
The Wigner-Eckart theorem is then used to separate the geometric part from the dynamic part of equation \ref{eq:signal}, resulting in relative intensities of the $F_3 $ lines that depend only on the dynamics of the preparation step and factors that depend on the electric quadrupole excitation polarization and the experimental geometry.
These relative intensities can be directly compared with the experimental data.
The calculation indicates that the decay geometry plays a minor role (less than 1 \%) in the relative intensities. 
On the other hand, the relative peak intensities strongly depend on the populations of the $5p_{3/2} $ magnetic sublevels produced in the preparation state.
Therefore, the electric quadrupole transition is at the same time a sensitive and non-perturbing probe of the preparation dynamics of the $5p_{3/2} \ M_2 $ magnetic sublevels.   

\begin{figure}
   \centering
       \includegraphics[width=0.8\linewidth]{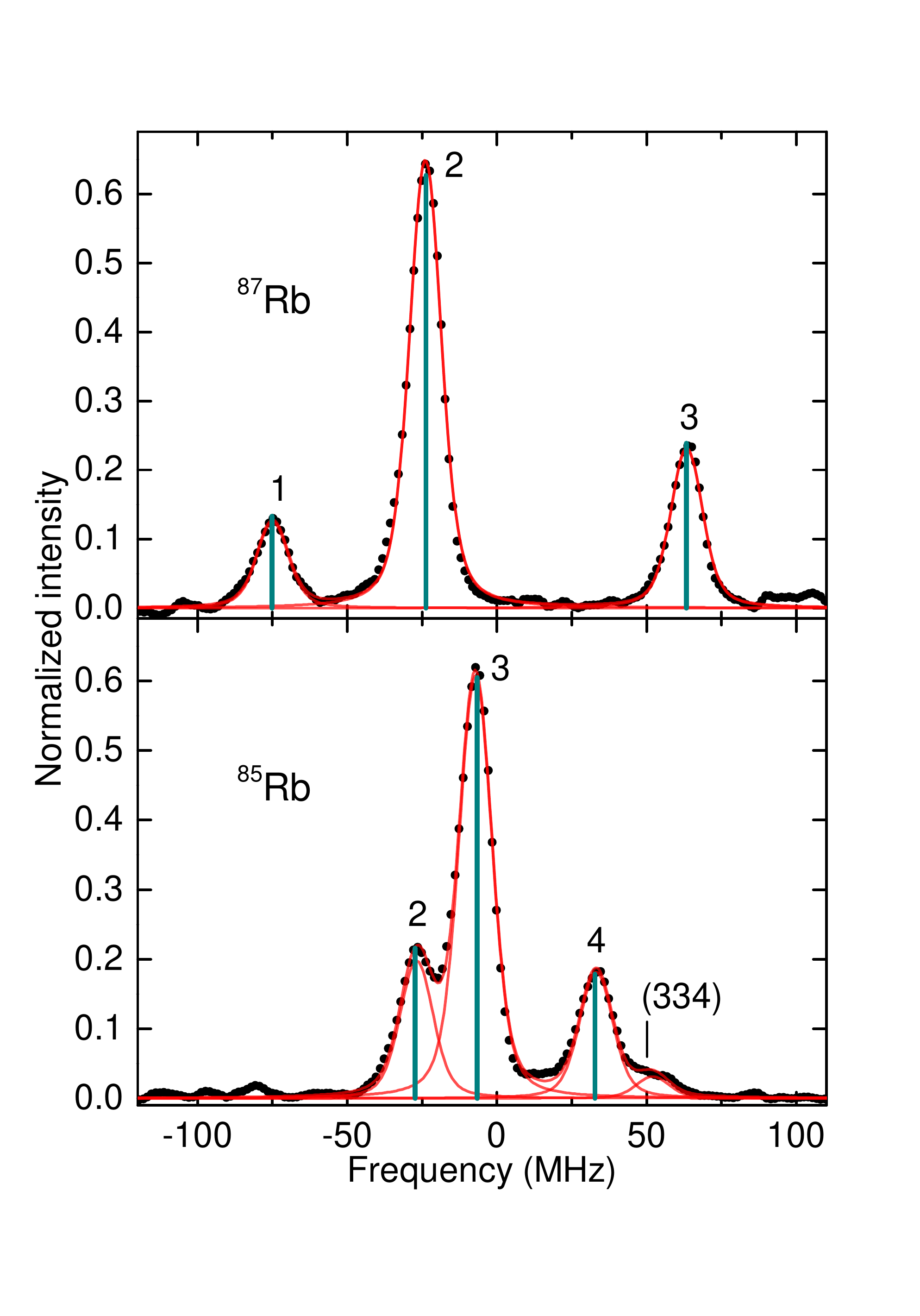}
       \caption{\label{fig:forbidden} Fluorescence emission spectra. Dots: experimental data,  continuous lines: result of fitting Voigt profiles to each line.The numeric labels indicate the $F $ value of the $6p_{3/2} $ hyperfine state. The velocity selected transition in $^{85} $Rb is indicated by the parenthesis with the F values of the excitation chain. The vertical bars give the position and calculated relative intensity of each hyperfine state. }
\end{figure}

Fig. \ref{fig:forbidden} shows typical spectra of the fluorescence signal recorded as the frequency of the $911 $ nm laser was scanned.
The original horizontal scale is the voltage applied to the ECDL2 piezo.
A coarse frequency equivalence is obtained with the Fabry-Perot interferometer. 
The least squares fits of independent Voigt profiles shown in the plot were performed for each spectrum.
The center and height of each peak was varied independently whilst the widths (Gaussian and Lorentzian) were the same for all peaks.
The differences between peak centers were then fit to the known $6p_{3/2} $ hyperfine splittings \cite{Arimondo1977}.
Finally, the zero in frequency was shifted to the center of gravity of the $6p_{3/2} $ hyperfine manifold common to both isotopes\cite{Arimondo1977}.
After this calibration the resulting total line width for both isotopes is $\Gamma = 12.9 \pm 0.2 $ MHz (FWHM).

For both isotopes we observe the expected three lines that result from the excitation sequence $5s_{1/2} F \rightarrow 5p_{3/2} F+1 \rightarrow 6p_{3/2} F_3 $, ($F_3 = F-1 $, $F $ and $F+1$) for zero velocity atoms.
Also, the splittings of the triplets correspond to the known frequency separation between the $6p_{3/2} F_3 $ hyperfine states of each isotope \cite{Arimondo1977}. 
However, other groups of atoms, with nonzero velocity projections, are also excited by the preparation laser. 
For these groups the Doppler shift of the counterpropagating $911 $ nm laser only partially compensates the Doppler shift of the preparation beam, and the dipole-forbidden transitions  appear at frequencies different to the ones obtained with the maximum $F $ preparation.
The strongest of these velocity-selected non-dipole transitions results from the $F \rightarrow F \rightarrow F+1 $ excitation chain ($2 \rightarrow 2 \rightarrow 3 $ in $^{87}$Rb and $3 \rightarrow 3 \rightarrow 4 $ in $^{85}$Rb).
In $^{85} $Rb there is a clear indication of a shoulder $\approx 19 $ MHz above the $F=4 $  peak, in good agreement with position of the velocity selected transition that is expected to appear $16.4 $ MHz above the $3 \rightarrow 3 \rightarrow 4 $ excitation.
No evidence of the corresponding $2 \rightarrow 2 \rightarrow 3 $ transition is found in the $^{87} $Rb spectrum. 
This transition is expected to occur $37 $ MHz above the $F_3 = 3 $ line in Fig. \ref{fig:forbidden}. 

The fit also gives information about the relative intensity of the hyperfine lines. 
For $^{85} $Rb the relative intensities are $20 $\%, $62 $\% and $18 $\%, while for $^{87} $Rb they are $12 $\%, $65 $\% and $23 $\%.
The calculated relative populations are $22 $\%, $60 $\% and $18 $\% for $^{85}$Rb and $13 $\%, $63 $\% and $24 $\% for $^{87}$Rb, in very good agreement with the measured values.
No variation of these ratios was found for values of the preparation laser power between $10 $ and $100 \ \mu$W.
This is in agreement with the calculation, that also predicts no significant change of the intensity ratios in this range of preparation laser intensities.

The intensity of the velocity selected peak in $^{85}$Rb ($3 \rightarrow 3 \rightarrow 4$) is $3.2 \% $ of the sum of intensities of the other three peaks.
In the calculation the electric quadrupole transition probability for this line is comparable to the ones in the zero velocity triplet. 
The reduction of its intensity is explained in terms of optical pumping effects, that effectively move the $F_2 = 3 $ population into the $F_1 = F-1=2 $ dark state.


In summary, direct evidence of the $5p_{3/2} \rightarrow 6p_{3/2} $ electric dipole forbidden excitation in atomic rubidium was presented.
The experiment was performed with continuous wave diode lasers and thermal atoms.
Efficient detection of the fluorescence that follows the Doppler-free optical-optical excitation allowed resolution of the $6p_{3/2} $ hyperfine structure.
Our results confirm that the $5p_{3/2} \rightarrow 6p_{3/2} $ excitation is the result of an electric quadrupole transition.
A calculation using a two-step excitation and one step decay is in very good agreement with the experiment.
This electric dipole forbidden transition is a very sensitive probe of the dynamics of the $5s \rightarrow 5p_{3/2} $ preparation step.

\begin{acknowledgments}
We thank J. Rangel for his help in the construction of the diode laser. This work was supported by DGAPA-UNAM, M\'exico, under projects PAPIIT Nos. IN116309, IN110812, and IA101012, by CONACyT, M\'exico, under Basic Research project No. 44986 and National Laboratory project LN260704.
\end{acknowledgments}


%

\end{document}